\documentclass[preprint,prd,aps,showpacs,showkeys,nofootinbib]{revtex4}
\usepackage{graphicx}
\begin{document}


\title{125 GeV Higgs boson decays in the $\mu$ from $\nu$ supersymmetric standard model}

\author{Hai-Bin Zhang$^{a,b,}$\footnote{Corresponding author. \\hbzhang@mail.dlut.edu.cn},
Tai-Fu Feng$^{a,b,}$\footnote{fengtf@hbu.edu.cn},
Fei Sun$^{a,b}$, Ke-Sheng Sun$^{a,b}$, Jian-Bin Chen$^{c}$,
Shu-Min Zhao$^{a,}$\footnote{smzhao@hbu.edu.cn}}

\affiliation{$^a$Department of Physics, Hebei University, Baoding, 071002, China\\
$^b$Department of Physics, Dalian University of Technology,
Dalian, 116024, China\\
$^c$Department of Physics, Taiyuan University of Technology,
Taiyuan, 030024, China}

\begin{abstract}
Recently the ATLAS and CMS Collaborations have reported significant events that are attributed to the neutral Higgs boson with mass around 125 GeV. In this work, we investigate the signals of the Higgs boson decay channels $h\rightarrow\gamma\gamma$, $h\rightarrow VV^*$ ($V=Z,W$), and $h\rightarrow f\bar{f}$ ($f=b,\tau$) in the $\mu$ from $\nu$ supersymmetric standard model ($\mu\nu$SSM). In the numerical results, we show the light stop and stau effects on the signal strengths for the 125 GeV Higgs boson decay channels in the $\mu\nu$SSM, which can account for the updated experimental data on Higgs.
\end{abstract}


\maketitle

\section{Introduction\label{sec1}}

As the simplest soft broken supersymmetry (SUSY) theory, the minimal supersymmetric standard model (MSSM)~\cite{MSSM} has attracted the attention of physicists for a long time. Furthermore, there is the SUSY extension of the standard model (SM), called the ``$\mu$ from $\nu$ supersymmetric standard model'' ($\mu\nu$SSM)~\cite{mnSSM,mnSSM1,mnSSM2}, which solves the $\mu$ problem~\cite{m-problem} of the MSSM through the lepton number  breaking couplings between the right-handed sneutrinos and the Higgses  $\epsilon _{ab}{\lambda _i}\hat \nu _i^c\hat H_d^a\hat H_u^b$ in the superpotential. Once the electroweak symmetry is broken (EWSB), the effective $\mu$ term $\epsilon _{ab} \mu \hat H_d^a\hat H_u^b$ is generated spontaneously through right-handed sneutrino vacuum expectation values (VEVs), $\mu  = {\lambda _i}\left\langle {\tilde \nu _i^c} \right\rangle$. Additionally, three tiny neutrino masses are generated at the tree level through a TeV scale seesaw mechanism~\cite{mnSSM,neutrino-mass}.

To understand the origin of the electroweak symmetry breaking and search for the neutral Higgs~\cite{Higgs-pred} predicted by the standard model and its various extensions is the main goal of the Large Hadron Collider (LHC). Recently the ATLAS and CMS Collaborations have reported significant excess events for a new boson, which is interpreted as the neutral Higgs with mass around 125 GeV at $5.0~\sigma$ level~\cite{ATLAS,CMS}. The CP properties and couplings of the particle are also being established~\cite{ATLAS1,CMS1}. This implies that the Higgs mechanism to break electroweak symmetry has a solid experimental cornerstone. In this paper, we investigate the 125 GeV Higgs decay channels $h\rightarrow\gamma\gamma$, $h\rightarrow VV^*$ ($V=Z,W$), and $h\rightarrow f\bar{f}$ ($f=b,\tau$) in the $\mu\nu$SSM. In the numerical analysis, we show the light stop and stau contributions to the signal strengths of the Higgs decay channels.

Our presentation is organized as follows. In Sec.~\ref{sec2}, we briefly summarize the main ingredients of the $\mu\nu$SSM by introducing its superpotential and the general soft SUSY-breaking terms, in particular discussing the Higgs sector. We present the decay widths and signal strengths for $h\rightarrow\gamma\gamma$, $h\rightarrow VV^*$ ($V=Z,W$) and $h\rightarrow f\bar{f}$ ($f=b,\tau$) in Sec.~\ref{sec3}. The numerical analysis are given in Sec.~\ref{sec4}, and Sec.~\ref{sec5} gives a summary.  The tedious formulae are collected in Appendixes~\ref{app-rad}--\ref{app-mini}.

\section{The $\mu\nu$SSM and the Higgs sector\label{sec2}}
Besides the superfields of the MSSM, the $\mu\nu$SSM introduces three singlet right-handed neutrino superfields $\hat{\nu}_i^c$. In addition to the MSSM Yukawa couplings for quarks and charged leptons, the superpotential of the $\mu\nu$SSM contains Yukawa couplings for neutrinos, two additional types of terms involving the Higgs doublet superfields $\hat H_d$ and $\hat H_u$, and the three right-handed neutrino superfields  $\hat{\nu}_i^c$,~\cite{mnSSM}
\begin{eqnarray}
&&W={\epsilon _{ab}}\left( {Y_{{u_{ij}}}}\hat H_u^b\hat Q_i^a\hat u_j^c + {Y_{{d_{ij}}}}\hat H_d^a\hat Q_i^b\hat d_j^c
+ {Y_{{e_{ij}}}}\hat H_d^a\hat L_i^b\hat e_j^c \right)  \nonumber\\
&&\hspace{0.95cm}
+ {\epsilon _{ab}}{Y_{{\nu _{ij}}}}\hat H_u^b\hat L_i^a\hat \nu _j^c -  {\epsilon _{ab}}{\lambda _i}\hat \nu _i^c\hat H_d^a\hat H_u^b + \frac{1}{3}{\kappa _{ijk}}\hat \nu _i^c\hat \nu _j^c\hat \nu _k^c ,
\label{eq-W}
\end{eqnarray}
where $\hat H_u^T = \Big( {\hat H_u^ + ,\hat H_u^0} \Big)$, $\hat H_d^T = \Big( {\hat H_d^0,\hat H_d^ - } \Big)$, $\hat Q_i^T = \Big( {{{\hat u}_i},{{\hat d}_i}} \Big)$, $\hat L_i^T = \Big( {{{\hat \nu}_i},{{\hat e}_i}} \Big)$ are $SU(2)$ doublet superfields, and $\hat u_i^c$, $\hat d_i^c$, and $\hat e_i^c$ represent the singlet up-type quark, down-type quark and charged lepton superfields, respectively.  In addition, $Y_{u,d,e,\nu}$, $\lambda$, and $\kappa$ are dimensionless matrices, a vector, and a totally symmetric tensor.  $a,b=1,2$ are SU(2) indices with antisymmetric tensor $\epsilon_{12}=1$, and $i,j,k=1,2,3$ are generation indices. The summation convention is implied on repeated indices in this paper.

In the superpotential, if the scalar potential is such that nonzero VEVs of the scalar components ($\tilde \nu _i^c$) of the singlet neutrino superfields $\hat{\nu}_i^c$ are induced, the effective bilinear terms $\epsilon _{ab} \varepsilon_i \hat H_u^b\hat L_i^a$ and $\epsilon _{ab} \mu \hat H_d^a\hat H_u^b$ are generated, with $\varepsilon_i= Y_{\nu _{ij}} \left\langle {\tilde \nu _j^c} \right\rangle$ and $\mu  = {\lambda _i}\left\langle {\tilde \nu _i^c} \right\rangle$,  once the electroweak symmetry is broken. The last term generates the effective Majorana masses for neutrinos at the electroweak scale, and the last two terms explicitly violate lepton number and R-parity. In SUSY extensions of the standard model, the R-parity of a particle is defined as $R = (-1)^{L+3B+2S}$~\cite{MSSM} and can be violated if either the baryon number ($B$) or lepton number ($L$) is not conserved, where $S$ denotes the spin of concerned component field. R-parity breaking implies that the lightest supersymmetric particle (LSP) is no longer stable. In this context, the neutralino or the sneutrino are no longer candidates for the dark matter. However, other SUSY particles such as the gravitino or the axino can still be used as candidates~\cite{mnSSM1}.

In the framework of supergravity-mediated supersymmetry breaking, the general soft SUSY-breaking terms of the $\mu\nu$SSM are given by
\begin{eqnarray}
&&- \mathcal{L}_{soft}=m_{{{\tilde Q}_{ij}}}^{\rm{2}}\tilde Q{_i^{a\ast}}\tilde Q_j^a
+ m_{\tilde u_{ij}^c}^{\rm{2}}\tilde u{_i^{c\ast}}\tilde u_j^c + m_{\tilde d_{ij}^c}^2\tilde d{_i^{c\ast}}\tilde d_j^c
+ m_{{{\tilde L}_{ij}}}^2\tilde L_i^{a\ast}\tilde L_j^a  \nonumber\\
&&\hspace{1.7cm} +  m_{\tilde e_{ij}^c}^2\tilde e{_i^{c\ast}}\tilde e_j^c + m_{{H_d}}^{\rm{2}} H_d^{a\ast} H_d^a
+ m_{{H_u}}^2H{_u^{a\ast}}H_u^a + m_{\tilde \nu_{ij}^c}^2\tilde \nu{_i^{c\ast}}\tilde \nu_j^c \nonumber\\
&&\hspace{1.7cm}  +  \epsilon_{ab}{\left[{{({A_u}{Y_u})}_{ij}}H_u^b\tilde Q_i^a\tilde u_j^c
+ {{({A_d}{Y_d})}_{ij}}H_d^a\tilde Q_i^b\tilde d_j^c + {{({A_e}{Y_e})}_{ij}}H_d^a\tilde L_i^b\tilde e_j^c + {\rm{H.c.}} \right]} \nonumber\\
&&\hspace{1.7cm}  + \left[ {\epsilon _{ab}}{{({A_\nu}{Y_\nu})}_{ij}}H_u^b\tilde L_i^a\tilde \nu_j^c
- {\epsilon _{ab}}{{({A_\lambda }\lambda )}_i}\tilde \nu_i^c H_d^a H_u^b
+ \frac{1}{3}{{({A_\kappa }\kappa )}_{ijk}}\tilde \nu_i^c\tilde \nu_j^c\tilde \nu_k^c + {\rm{H.c.}} \right] \nonumber\\
&&\hspace{1.7cm}  -  \frac{1}{2}\left({M_3}{{\tilde \lambda }_3}{{\tilde \lambda }_3}
+ {M_2}{{\tilde \lambda }_2}{{\tilde \lambda }_2} + {M_1}{{\tilde \lambda }_1}{{\tilde \lambda }_1} + {\rm{H.c.}} \right).
\end{eqnarray}
Here, the first two lines contain mass squared terms of squarks, sleptons, and Higgses. The next two lines consist of the trilinear scalar couplings. In the last line, $M_3$, $M_2$, and $M_1$ denote Majorana masses corresponding to $SU(3)$, $SU(2)$, and $U(1)$ gauginos $\hat{\lambda}_3$, $\hat{\lambda}_2$, and $\hat{\lambda}_1$, respectively. In addition to the terms from $\mathcal{L}_{soft}$, the tree-level scalar potential receives the usual $D$- and $F$-term contributions~\cite{mnSSM1}.

Once the electroweak symmetry is spontaneously broken, the neutral scalars develop in general the VEVs:
\begin{eqnarray}
\langle H_d^0 \rangle = \upsilon_d , \qquad \langle H_u^0 \rangle = \upsilon_u , \qquad
\langle \tilde \nu_i \rangle = \upsilon_{\nu_i} , \qquad \langle \tilde \nu_i^c \rangle = \upsilon_{\nu_i^c} .
\end{eqnarray}
One can define the neutral scalars as
\begin{eqnarray}
&&H_d^0=\frac{h_d + i P_d}{\sqrt{2}} + \upsilon_d, \qquad\; \tilde \nu_i = \frac{(\tilde \nu_i)^\Re + i (\tilde \nu_i)^\Im}{\sqrt{2}} + \upsilon_{\nu_i},  \nonumber\\
&&H_u^0=\frac{h_u + i P_u}{\sqrt{2}} + \upsilon_u, \qquad \tilde \nu_i^c = \frac{(\tilde \nu_i^c)^\Re + i (\tilde \nu_i^c)^\Im}{\sqrt{2}} + \upsilon_{\nu_i^c},
\end{eqnarray}
and
\begin{eqnarray}
\tan\beta={\upsilon_u\over\sqrt{\upsilon_d^2+\upsilon_{\nu_i}\upsilon_{\nu_i}}}.
\end{eqnarray}

For simplicity, we will assume that all parameters in the potential are real. The CP-odd neutral scalar and charged scalar mass matrices can isolate massless unphysical Goldstone bosons $G^0$ and $G^{\pm}$, which can be written as \cite{Zhang,Zhang1}
\begin{eqnarray}
&&G^0 = {1 \over \upsilon_{_{\rm{EW}}}} \Big(\upsilon_d {P_d}-\upsilon_u{P_u}-\upsilon_{\nu_i}{(\tilde \nu_i)^\Im}\Big),\nonumber\\
&&G^{\pm} = {1 \over \upsilon_{_{\rm{EW}}}} \Big(\upsilon_d H_d^{\pm} - \upsilon_u {H_u^{\pm}}-\upsilon_{\nu_i}\tilde e_{L_i}^{\pm}\Big),
\end{eqnarray}
through an $8\times8$ unitary matrix $Z_H$,
\begin{eqnarray}
&&Z_H=\left(\begin{array}{*{20}{c}}
{\upsilon_d\over\upsilon_{_{\rm EW}}}&{\upsilon_u\over\upsilon_{_{\rm SM}}}&
\Big({\upsilon_d\upsilon_{\nu_i}\over\upsilon_{_{\rm SM}}\upsilon_{_{\rm EW}}}\Big)_{{1\times3}}\\ [6pt]
-{\upsilon_u\over\upsilon_{_{\rm EW}}}&{\upsilon_d\over\upsilon_{_{\rm SM}}}&
\Big(-{\upsilon_u\upsilon_{\nu_i}\over\upsilon_{_{\rm SM}}\upsilon_{_{\rm EW}}}\Big)_{{1\times3}}\\ [6pt]
\Big(-{\upsilon_{\nu_i}\over\upsilon_{_{\rm EW}}}\Big)_{{3\times1}}&0_{{3\times1}}&\Big({\upsilon_{_{\rm SM}}\over\upsilon_{_{\rm EW}}}
\delta_{ij}+\varepsilon_{ijk}{\upsilon_{\nu_k}\over\upsilon_{_{\rm EW}}}\Big)_{{3\times3}}
\end{array}\right)\bigoplus1_{{3\times3}},
\end{eqnarray}
where
\begin{eqnarray}
\upsilon_{_{\rm{SM}}}=\sqrt{\upsilon_d^2+\upsilon_u^2},\qquad \upsilon_{_{\rm{EW}}}=\sqrt{\upsilon_d^2+\upsilon_u^2+\upsilon_{\nu_i}\upsilon_{\nu_i}}.
\end{eqnarray}
In the physical gauge, the Goldstone bosons $G^0$ and $G^{\pm}$ are, respectively, eaten by the $Z$ boson and $W$ boson and disappear from the Lagrangian.
The masses squared of the $Z$ boson and $W$ boson are
\begin{eqnarray}
m_Z^2={e^2\over {2s_{W}^2 c_{W}^2}}\upsilon_{_{\rm{EW}}}^2, \qquad m_W^2={e^2\over2s_{W}^2}\upsilon_{_{\rm{EW}}}^2.
\end{eqnarray}
Here $e$ is the electromagnetic coupling constant, $s_{W}=\sin\theta_{W}$ and $c_{W}=\cos\theta_{W}$, with $\theta_{W}$ denoting the Weinberg angle, respectively.

In the $\mu\nu$SSM, the VEVs of left- and right-handed sneutrinos lead to mixing of the neutral components of the Higgs doublets with the sneutrinos producing an $8\times8$ CP-even  neutral scalar mass matrix. However, if the off-diagonal mixing terms of the CP-even neutral scalar mass matrix are small enough than the diagonal terms, the contribution of the off-diagonal mixing terms to the diagonal doubletlike Higgs masses is small and can be neglected. Actually, we will use this mechanism in our calculation.

Considering radiative corrections, the mass squared matrix for the neutral Higgs doublets in the basis $(h_d,\;h_u)$ is written as
\begin{eqnarray}
{\cal M}^2=
\left(\begin{array}{*{20}{c}}
M_{h_d h_d}^2+\Delta_{11} & M_{h_d h_u}^2+\Delta_{12}\\  [6pt]
M_{h_d h_u}^2+\Delta_{12} & M_{h_u h_u}^2+\Delta_{22}
\end{array}\right),
\label{eq-M}
\end{eqnarray}
with
\begin{eqnarray}
&&M_{h_d h_u}^2\simeq -\Big[m_A^2+\Big(1-4\lambda_i \lambda_i s_{W}^2 c_{W}^2/e^2 \Big)m_Z^2\Big] \sin \beta \cos \beta, \nonumber\\
&&M_{h_d h_d}^2\simeq m_A^2 \sin^2 \beta+m_Z^2 \cos^2 \beta, \nonumber\\
&&M_{h_u h_u}^2\simeq m_A^2 \cos^2 \beta+m_Z^2 \sin^2 \beta,
\end{eqnarray}
and the neutral pseudoscalar mass squared is
\begin{eqnarray}
m_A^2\simeq 2\Big[(A_\lambda \lambda)_i \upsilon_{\nu_i^c}+ {\lambda _k}{\kappa _{ijk}}\upsilon_{\nu_i^c} \upsilon_{\nu_j^c}\Big] /{\sin2\beta}.
\end{eqnarray}
Compared with the MSSM, $M_{h_d h_u}^2$ gets an additional term $(4\lambda_i \lambda_i s_{W}^2 c_{W}^2/e^2 )m_Z^2 \sin \beta \cos \beta$, which gives a new contribution to the light doubletlike Higgs mass. In Eq.~(\ref{eq-M}), the concrete expressions for radiative corrections $\Delta_{11}$, $\Delta_{12}$ and $\Delta_{22}$ can be found in Appendix~\ref{app-rad}.
Besides the superfields of the MSSM, the $\mu\nu$SSM introduces right-handed neutrino superfields. Nevertheless, the loop effects of right-handed neutrino/sneutrino on the light doubletlike Higgs boson mass can be neglected, due to small neutrino Yukawa couplings $Y_{\nu_i} \sim \mathcal{O}(10^{-7})$ and left-handed sneutrino VEVs $\upsilon_{\nu_i} \sim \mathcal{O}(10^{-4}{\rm{GeV}})$. Through the numerical computation, we can ignore the radiative corrections from $b$ quark, $\tau$ lepton, and their supersymmetric partners, when $\tan \beta$ is small. The main radiative corrections on the light doubletlike Higgs boson mass come from the top quark and its supersymmetric partners, similarly to the MSSM. By the $2\times2$ unitary matrix $U_h$,
\begin{eqnarray}
U_h=
\left(\begin{array}{*{20}{c}}
-\sin \alpha & \cos \alpha\\
\cos \alpha & \sin \alpha
\end{array}\right),
\end{eqnarray}
the mass squared matrix ${\cal M}^2$ which contains the radiative corrections can be diagonalized:
\begin{eqnarray}
U_h^T {\cal M}^2 U_h = diag\Big(m_h^2,m_H^2\Big).
\end{eqnarray}
Here the neutral doubletlike Higgs mass squared eigenvalues $m_{h(H)}^2$ can be derived~\cite{Higgs-Carena},
\begin{eqnarray}
m_{h(H)}^2={1\over 2}\Big({\rm{Tr}}{\cal M}^2 \mp\sqrt{({{\rm{Tr}}{\cal M}^2})^2-4{\rm{Det}}{\cal M}^2}\Big),
\end{eqnarray}
where ${\rm{Tr}}{\cal M}^2={\cal M}_{11}^2+{\cal M}_{22}^2$, ${\rm{Det}}{\cal M}^2 = {\cal M}_{11}^2{\cal M}_{22}^2-({\cal M}_{12}^2)^2$.
The mixing angle $\alpha$ can be determined by~\cite{alpha}
\begin{eqnarray}
&&\sin 2\alpha =\frac{2{\cal M}^2_{12}}{\sqrt{({{\rm{Tr}}{\cal M}^2})^2-4{\rm{Det}}{\cal M}^2}},\nonumber\\
&&\cos 2\alpha =\frac{{\cal M}^2_{11}-{\cal M}^2_{22}}{\sqrt{({{\rm{Tr}}{\cal M}^2})^2-4{\rm{Det}}{\cal M}^2}},
\end{eqnarray}
which reduce to $-\sin 2\beta$ and $-\cos 2\beta$, respectively, in the large $m_A$ limit. The convention is that $\pi/4\leq\beta<\pi/2$ for $\tan \beta\geq1$, while $-\pi/2<\alpha<0$. In the large $m_A$ limit, $\alpha=-\pi/2+\beta$.

One most stringent constraint on parameter space of the $\mu\nu$SSM is that the mass squared matrix should produce an eigenvalue around $(125\:{\rm GeV})^2$
as mass squared of the light doubletlike Higgs. The combination of the ATLAS~\cite{ATLAS} and CMS~\cite{CMS} for the neutral Higgs mass gives~\cite{mh-AC}
\begin{eqnarray}
m_{{h}}=125.7\pm0.4\;{\rm GeV}.
\label{M-h}
\end{eqnarray}
This fact constrains parameter space of the $\mu\nu$SSM stringently.

\section{The 125 GeV Higgs decays\label{sec3}}
At the LHC, the Higgs can be mainly produced by the gluon fusion. In the SM, the leading-order (LO) contributions originate from the one-loop diagrams involving virtual top quark.  The cross section for this process is known to the next-to-next-to-leading order (NNLO)~\cite{NNLO} which can enhance the LO result by 80\%-100\%. Beyond the SM, any new particle which strongly couples with the Higgs can modify the cross section of the process significantly. In the new physics (NP), the LO decay width of $h\rightarrow gg$ process is given as (see Ref.~\cite{Gamma1} and references therein)
\begin{eqnarray}
\Gamma_{{\rm{NP}}}(h\rightarrow gg)={G_{F}\alpha_s^2m_{h}^3\over64\sqrt{2}\pi^3}
\Big|\sum\limits_q g_{{hqq}}A_{1/2}(x_q)
+\sum\limits_{\tilde q} g_{{h\tilde{q}\tilde{q}}}{m_Z^2\over m_{{\tilde q}}^2}A_{0}(x_{{\tilde{q}}})\Big|^2,
\label{hgg}
\end{eqnarray}
with $x_a=m_{{h}}^2/(4m_a^2)$, $q=t,\:b$,
and $\tilde{q}=U_I^+,\:D_I^- \: (I=1,\ldots,6)$.
The concrete expressions of $g_{{htt}},\:g_{{hbb}},\:g_{{h U_I^- U_I^+}}
,\:g_{{h D_I^+ D_I^-}}$ are formulated as
\begin{eqnarray}
&&g_{htt}={\cos \alpha\over \sin \beta} ,   \nonumber\\
&&g_{hbb}=-{\sin \alpha\over \cos \beta}  \sqrt{1+\sum\limits_{i=1}^3 {\upsilon_{\nu_i}^2 \over \upsilon_d^2} },   \nonumber\\
&&g_{{h U_I^- U_I^+}}=-{ \upsilon_{_{\rm{EW}}}\over 2 m_{Z}^2} C_{1 II}^{U^\pm} \quad (I=1,\ldots,6),
\nonumber\\
&&g_{{h D_I^+ D_I^-}}=-{ \upsilon_{_{\rm{EW}}}\over 2 m_{Z}^2} C_{1 II}^{D^\pm} \quad (I=1,\ldots,6),
\end{eqnarray}
where the concrete expressions for $C_{1 II}^{U^\pm},\: C_{1 II}^{D^\pm}$ can be
found in Appendix~\ref{app-coupling}. The form factors $A_0,\: A_{1/2}$ (and $A_1$ below) are defined in  Appendix~\ref{app-form}. In Eq.~(\ref{hgg}), the contributions of squarks have the damped loop factors ${m_Z^2 / m_{{\tilde q}}^2}$. Thus, contrary to the case of SM quarks, the contributions of squarks become very small for high masses, and the squarks decouple completely from the gluonic Higgs couplings if they are very heavy.

The decay width of the Higgs to diphoton decay at LO in the SM is derived from the
one-loop diagrams which contain virtual top quark or virtual $W$ boson. In the NP, the third- generation fermions ($f=$ $t$, $b$, $\tau$) and $W$ boson
together with the supersymmetric partners give the contributions to the LO decay width for the Higgs to diphoton decay, which can be written by
\begin{eqnarray}
&&\Gamma_{{\rm{NP}}}(h\rightarrow\gamma\gamma)={G_{F}\alpha^2m_{{h}}^3\over128\sqrt{2}\pi^3}
\Big|\sum\limits_f N_c Q_{f}^2 g_{{hff}} A_{1/2}(x_f)+g_{{hWW}}A_1(x_{W})
\nonumber\\
&&\hspace{3.0cm}
+\sum\limits_{\alpha=2}^8 g_{{h S_\alpha^+ S_\alpha^-}}{m_{Z}^2\over m_{S_\alpha^\pm}^2}A_0(x_{S_\alpha^\pm})
+\sum\limits_{i=1}^2 g_{{h\chi_i^+\chi_i^-}}{m_{W}\over m_{\chi_i^\pm}}A_{1/2}(x_{\chi_i^\pm})
\nonumber\\
&&\hspace{3.0cm}
+\sum\limits_{\tilde q} N_c Q_{f}^2 g_{{h\tilde{q}\tilde{q}}}{m_{Z}^2\over m_{{\tilde q}}^2}A_{0}(x_{{\tilde{q}}})\Big|^2,
\end{eqnarray}
and the expressions of $g_{{h\tau\tau}}$, $g_{{hWW}}$, $g_{{h S_\alpha^+ S_\alpha^-}}$, $g_{{h\chi_i^+\chi_i^-}}$ are
\begin{eqnarray}
&&g_{h\tau\tau}\simeq-{\sin \alpha\over \cos \beta}  \sqrt{1+\sum\limits_{i=1}^3 {\upsilon_{\nu_i}^2 \over \upsilon_d^2} },   \nonumber\\
&&g_{hWW}\simeq \sin ( \beta-\alpha) ,   \nonumber\\
&&g_{{h S_\rho^+ S_\rho^-}}=-{ \upsilon_{_{\rm{EW}}}\over 2 m_{Z}^2} C_{1 \rho\rho}^{S^\pm} \quad (\rho=2,\ldots,8),
\nonumber\\
&&g_{{h\chi_i^+\chi_i^-}}=-{2\over { e} } \Re\Big[C_{1 ii}^{\chi^\pm}\Big] \quad (i=1,2),
\end{eqnarray}
where the concrete expressions of $C_{1 \rho\rho}^{S^\pm},\: C_{1 ii}^{\chi^\pm}$ can be found in Appendix~\ref{app-coupling}. Here, if supersymmetric particles are heavy, the contributions of supersymmetric particles will become small. And then, the main contributions of the Higgs to diphoton decay width at LO is derived from top quark, bottom quark, and $W$ boson.

The light doubletlike Higgs with $125\:{\rm GeV}$ mass can decay through the channels  $h\rightarrow WW^*$, $h\rightarrow ZZ^*$ where $V^*$ ($V=Z,W$) denoting the off-shell electroweak gauge bosons. Summing over all modes available to the $W^*$ or $Z^*$, the decay widths are given by~\cite{Keung1}
\begin{eqnarray}
&&\Gamma_{{\rm{NP}}}(h\rightarrow ZZ^*)={e^4m_h\over2048\pi^3s_W^4c_W^4}|g_{{hZZ}}|^2
\Big(7-{40\over3}s_W^2+{160\over9}s_W^4\Big)F({m_Z\over m_h}), \nonumber\\
&&\Gamma_{{\rm{NP}}}(h\rightarrow WW^*)={3e^4m_{{h}}\over512\pi^3s_{W}^4}|g_{{hWW}}|^2
F({m_{W}\over m_h}),
\end{eqnarray}
with $g_{{hZZ}}=g_{{hWW}}$ and the form factor $F(x)$ is formulated in Appendix~\ref{app-form}.
The partial decay width of the $125\:{\rm GeV}$ neutral Higgs into fermion pairs is
given in the Born approximation by~\cite{hff}
\begin{eqnarray}
&&\Gamma_{{\rm{NP}}}(h\rightarrow f\bar{f})=N_c {G_{F} m_{f}^2 m_h \over 4\sqrt{2}\pi}|g_{{hff}}|^2 (1-4m_f^2/m_h^2)^{3/2} \quad (f=b,\tau),
\end{eqnarray}
with $g_{{hbb}}\simeq g_{{h\tau\tau}}$.

Normalized to the SM expectation, the signal strengths for the Higgs decay channels are quantified by the ratios~\cite{ratios}
\begin{eqnarray}
&&\mu_{\gamma\gamma,VV^*}^{{\rm{ggF}}}={ \sigma_{{\rm{NP}}}({\rm{ggF}})\over
\sigma_{{\rm{SM}}}({\rm{ggF}})}\:{{\rm{BR}}_{{\rm{NP}}}(h\rightarrow\gamma\gamma,VV^*)\over
{\rm{BR}}_{{\rm{SM}}}(h\rightarrow\gamma\gamma,VV^*)}   \qquad (V=Z,W), \nonumber\\
&&\mu_{f\bar{f}}^{{\rm{VBF}}}={ \sigma_{{\rm{NP}}}({\rm{VBF}})\over
\sigma_{{\rm{SM}}}({\rm{VBF}})}\:{{\rm{BR}}_{{\rm{NP}}}(h\rightarrow{f\bar{f}})\over
{\rm{BR}}_{{\rm{SM}}}(h\rightarrow{f\bar{f}})}  \qquad (f=b,\tau),
\label{eq-ratios}
\end{eqnarray}
where ggF and VBF stand for gluon-gluon fusion and vector boson fusion, respectively. Normalized to the SM values, one can evaluate the Higgs production cross sections
\begin{eqnarray}
&&{\sigma_{{\rm{NP}}}({\rm{ggF}})\over
\sigma_{{\rm{SM}}}({\rm{ggF}})} \approx {\Gamma_{{\rm{NP}}}(h\rightarrow gg) \over
\Gamma_{{\rm{SM}}}(h\rightarrow gg)} = {\Gamma_{{\rm{NP}}}^h\over
\Gamma_{{\rm{SM}}}^h}\: {\Gamma_{{\rm{NP}}}(h\rightarrow gg)/\Gamma_{{\rm{NP}}}^h \over
\Gamma_{{\rm{SM}}}(h\rightarrow gg)/\Gamma_{{\rm{SM}}}^h}\nonumber\\
&&\qquad\qquad\;\;={\Gamma_{{\rm{NP}}}^h\over
\Gamma_{{\rm{SM}}}^h}\: {{\rm{BR}}_{{\rm{NP}}}(h\rightarrow gg)\over
{\rm{BR}}_{{\rm{SM}}}(h\rightarrow gg)},\nonumber\\
&&{\sigma_{{\rm{NP}}}({\rm{VBF}})\over
\sigma_{{\rm{SM}}}({\rm{VBF}})} \approx {\Gamma_{{\rm{NP}}}(h\rightarrow{VV^*}) \over
\Gamma_{{\rm{SM}}}(h\rightarrow{VV^*})} ={\Gamma_{{\rm{NP}}}^h\over
\Gamma_{{\rm{SM}}}^h}\: {\Gamma_{{\rm{NP}}}(h\rightarrow{VV^*})/\Gamma_{{\rm{NP}}}^h \over
\Gamma_{{\rm{SM}}}(h\rightarrow{VV^*})/\Gamma_{{\rm{SM}}}^h}\nonumber\\
&&\qquad\qquad\quad ={\Gamma_{{\rm{NP}}}^h\over
\Gamma_{{\rm{SM}}}^h}\: {{\rm{BR}}_{{\rm{NP}}}(h\rightarrow{VV^*})\over
{\rm{BR}}_{{\rm{SM}}}(h\rightarrow{VV^*})},
\label{eq-cross}
\end{eqnarray}
with the 125 GeV Higgs total decay width for the NP
\begin{eqnarray}
&&\Gamma_{{\rm{NP}}}^h=\sum\limits_{f=b,\tau,c,s} \Gamma_{{\rm{NP}}}(h\rightarrow f\bar{f})+ \sum\limits_{V=Z,W} \Gamma_{{\rm{NP}}}(h\rightarrow VV^*) \nonumber\\
&&\qquad\quad +\: \Gamma_{{\rm{NP}}}(h\rightarrow gg) +\Gamma_{{\rm{NP}}}(h\rightarrow \gamma\gamma),
\end{eqnarray}
where we have neglected the contributions from the rare or invisible decays, and $\Gamma_{{\rm{SM}}}^h$ denotes the SM Higgs total decay width. Through Eq.~(\ref{eq-ratios}) and Eq.~(\ref{eq-cross}), we can quantify the signal strengths for the Higgs decay channels in the $\mu\nu$SSM
\begin{eqnarray}
&&\mu_{\gamma\gamma}^{{\rm{ggF}}}\approx {\Gamma_{{\rm{NP}}}(h\rightarrow gg) \over
\Gamma_{{\rm{SM}}}(h\rightarrow gg)} \:{\Gamma_{{\rm{NP}}}(h\rightarrow\gamma\gamma)/\Gamma_{{\rm{NP}}}^h\over
\Gamma_{{\rm{SM}}}(h\rightarrow\gamma\gamma)/\Gamma_{{\rm{SM}}}^h} \nonumber\\
&&\qquad\;\, ={\Gamma_{{\rm{SM}}}^h\over \Gamma_{{\rm{NP}}}^h}\:{\Gamma_{{\rm{NP}}}(h\rightarrow gg) \over
\Gamma_{{\rm{SM}}}(h\rightarrow gg)}\:{\Gamma_{{\rm{NP}}}(h\rightarrow\gamma\gamma)\over
\Gamma_{{\rm{SM}}}(h\rightarrow\gamma\gamma)}, \nonumber\\
&&\mu_{VV^*}^{{\rm{ggF}}}\approx {\Gamma_{{\rm{NP}}}(h\rightarrow gg)\over
\Gamma_{{\rm{SM}}}(h\rightarrow gg)} \:{\Gamma_{{\rm{NP}}}(h\rightarrow VV^*) /\Gamma_{{\rm{NP}}}^h \over \Gamma_{{\rm{SM}}}(h\rightarrow VV^*) /\Gamma_{{\rm{SM}}}^h} \nonumber\\
&&\qquad\;\, ={\Gamma_{{\rm{SM}}}^h\over \Gamma_{{\rm{NP}}}^h}\:{\Gamma_{{\rm{NP}}}(h\rightarrow gg) \over \Gamma_{{\rm{SM}}}(h\rightarrow gg)} \: |g_{{hVV}}|^2, \nonumber\\
&&\mu_{f\bar{f}}^{{\rm{VBF}}} \approx {\Gamma_{{\rm{NP}}}(h\rightarrow{VV^*}) \over
\Gamma_{{\rm{SM}}}(h\rightarrow{VV^*})} \:{\Gamma_{{\rm{NP}}}(h\rightarrow f\bar{f}) /\Gamma_{{\rm{NP}}}^h \over \Gamma_{{\rm{SM}}}(h\rightarrow f\bar{f}) /\Gamma_{{\rm{SM}}}^h} \nonumber\\
&&\qquad\;\, ={\Gamma_{{\rm{SM}}}^h\over \Gamma_{{\rm{NP}}}^h}\: |g_{{hVV}}|^2 \: |g_{{hff}}|^2\qquad (V=Z,W;f=b,\tau),
\label{signals}
\end{eqnarray}
with  ${\Gamma_{{\rm{NP}}}(h\rightarrow{VV^*}) \over
\Gamma_{{\rm{SM}}}(h\rightarrow{VV^*})} = |g_{{hVV}}|^2=|g_{{hZZ}}|^2=|g_{{hWW}}|^2$ and ${\Gamma_{{\rm{NP}}}(h\rightarrow f\bar{f}) \over \Gamma_{{\rm{SM}}}(h\rightarrow f\bar{f}) } =|g_{{hff}}|^2=|g_{{hbb}}|^2\simeq |g_{{h\tau\tau}}|^2$. Therefore, we could just analyze the signal strengths $\mu_{\gamma\gamma}^{{\rm{ggF}}}$, $\mu_{VV^*}^{{\rm{ggF}}}$ and $\mu_{f\bar{f}}^{{\rm{VBF}}}$ in the following.

\begin{table*}
\begin{tabular*}{\textwidth}{@{\extracolsep{\fill}}lllll@{}}
\hline
Signal & Value from ATLAS & Value from CMS & Weighted average \\
\hline
$\mu_{\gamma\gamma}^{{\rm{ggF}}}$ & $1.6_{-0.36}^{+0.42}$~\cite{ATLAS7} & $0.52\pm0.5$~\cite{CMS4} & $1.19\pm0.31$ \\
$\mu_{ZZ^*}^{{\rm{ggF}}}$ & $1.8_{-0.5}^{+0.8}$~\cite{ATLAS8} & $0.9_{-0.4}^{+0.5}$~\cite{CMS5} & $1.18\pm0.37$  \\
$\mu_{WW^*}^{{\rm{ggF}}}$ & $0.82\pm0.36$~\cite{ATLAS9} & $0.76\pm0.21$~\cite{CMS6} & $0.78\pm0.18$  \\
\hline
\end{tabular*}
\caption{Experimental values for the Higgs decay rates.}
\label{tab1}
\end{table*}

The latest LHC measurements of the Higgs decay rates are summarized in Table~\ref{tab1}, where we also compute the weighted averages for the signal strengths $\mu_{\gamma\gamma,VV^*}^{{\rm{ggF}}}$ ($V=Z,W$) from ATLAS and CMS. When the errors are asymmetric, we average them in quadrature. Note that for the signal strengths $\mu_{\gamma\gamma,ZZ^*}^{{\rm{ggF}}}$, the average from ATLAS and CMS just is used as a guideline, but this should be taken with some care as the two experiments have quite different central values. As the signal strengths $\mu_{ZZ^*}^{{\rm{ggF}}}$ and $\mu_{WW^*}^{{\rm{ggF}}}$ depend on the same couplings, we combine them and give a weighted average
\begin{eqnarray}
\mu_{VV^*}^{{\rm{ggF}}}=0.86\pm0.16,
\end{eqnarray}
to constrain the numerical evolution of $\mu_{VV^*}^{{\rm{ggF}}}$ in the following. Since the measured rates for the channels $h\rightarrow f\bar{f}$ still have large experimental error at now~\cite{CDFD0,CMS7,ATLAS10}, here we will not consider their experimental values to constrain the channels $h\rightarrow f\bar{f}$ in the $\mu\nu$SSM.

\section{Numerical analysis\label{sec4}}
There are many free parameters in the SUSY extensions of the SM. In order to obtain a transparent numerical results, we make some assumptions on parameter space for the $\mu\nu{\rm SSM}$ before performing the numerical calculation. In the following, we make the minimal flavor violation (MFV) assumption
\begin{eqnarray}
&&\;\,{\kappa _{ijk}} = \kappa {\delta _{ij}}{\delta _{jk}}, \;\;
{({A_\kappa }\kappa )_{ijk}} =
{A_\kappa }\kappa {\delta _{ij}}{\delta _{jk}}, \quad\;\,
\lambda _i = \lambda , \qquad\quad\;\:
{{\rm{(}}{A_\lambda }\lambda {\rm{)}}_i} = {A_\lambda }\lambda,\nonumber\\
&&\;\,{Y_{{u _{ij}}}} = {Y_{{u _i}}}{\delta _{ij}},\quad
 (A_u Y_u)_{ij}={A_{u_i}}{Y_{{u_i}}}{\delta _{ij}},\quad
{Y_{{\nu _{ij}}}} = {Y_{{\nu _i}}}{\delta _{ij}},\quad (A_\nu Y_\nu)_{ij}={a_{{\nu_i}}}{\delta _{ij}},\nonumber\\
&&\;\,{Y_{{d_{ij}}}} = {Y_{{d_i}}}{\delta _{ij}},\quad\:
(A_d Y_d)_{ij}={A_{d_i}}{Y_{{d_i}}}{\delta _{ij}},\;\;\;\:
{Y_{{e_{ij}}}} = {Y_{{e_i}}}{\delta _{ij}},\quad\,
{({A_e}{Y_e})_{ij}} = {A_{e_i}}{Y_{{e_i}}}{\delta _{ij}},
\nonumber\\
&&m_{{{\tilde L}_{ij}}}^2 = m_{{\tilde L}_i}^2{\delta _{ij}},\qquad\;\;
m_{\tilde \nu_{ij}^c}^2 = m_{{{\tilde \nu_i}^c}}^2{\delta _{ij}},\quad\;\;\:
m_{\tilde e_{ij}^c}^2 = m_{{{\tilde e_i}^c}}^2{\delta _{ij}},\qquad\quad\, \upsilon_{\nu_i^c}=\upsilon_{\nu^c}, \nonumber\\
&&m_{\tilde Q_{ij}}^2 = m_{{{\tilde Q}_i}}^2{\delta _{ij}}, \qquad\;\,
m_{\tilde u_{ij}^c}^2 = m_{{{\tilde u}_i^c}}^2{\delta _{ij}}, \quad\;\;\,
m_{\tilde d_{ij}^c}^2 = m_{{{\tilde d}_i^c}}^2{\delta _{ij}},
\label{MFV}
\end{eqnarray}
where $i,\;j,\;k =1,\;2,\;3 $.

Restrained by the quark and lepton masses, we have
\begin{eqnarray}
{Y_{{u_i}}} = \frac{{{m_{{u_i}}}}}{{{\upsilon_u}}},\qquad {Y_{{d_i}}} = \frac{{{m_{{d_i}}}}}{{{\upsilon_d}}},\qquad {Y_{{e_i}}} = \frac{{{m_{{l_i}}}}}{{{\upsilon_d}}},
\end{eqnarray}
where $m_{u_i}$, $m_{d_i}$ and $m_{l_i}$ are the up-quark, down-quark and charged lepton masses, respectively, and the values are taken from the PDG~\cite{PDG}. For the masses of bino and wino, we will imply the approximate GUT relation $M_1=\frac{\alpha_1^2}{\alpha_2^2}M_2\approx 0.5 M_2$.
In  Appendix~\ref{app-mini}, the tree-level tadpoles, Eqs.~(\ref{potential-1})--(\ref{potential-4}), are set to be zero to minimize the potential. In this way, the soft masses $m_{\tilde H_d}^2$, $m_{\tilde H_u}^2$ and $m_{\tilde \nu_i^c}^2$ can be derived. Simultaneously, ignoring the terms of the second order in $Y_{\nu}$ and assuming $(\upsilon_{\nu_i}^2+\upsilon_d^2-\upsilon_u^2)\approx (\upsilon_d^2-\upsilon_u^2)$, one can solve the minimization conditions of the neutral scalar potential with respect to $\upsilon_{\nu_i}\:(i=1,2,3)$ as~\cite{neutrino-mass}:
\begin{eqnarray}
\upsilon_{\nu_i}=\frac{\lambda \upsilon_d (\upsilon_u^2+\upsilon_{\nu^c}^2) - \kappa \upsilon_u \upsilon_{\nu^c}^2}{m_{{{\tilde L}_i}}^2 +{G^2\over 4} (\upsilon_d^2-\upsilon_u^2)} Y_{\nu_i} -\frac{\upsilon_u \upsilon_{\nu^c}}{m_{{\tilde L}_i}^2 +{G^2\over 4} (\upsilon_d^2-\upsilon_u^2)}a_{\nu_i},
\label{eq-min}
\end{eqnarray}
where $G^2=g_1^2+g_2^2$ and $g_1 c_{_W} =g_2 s_{_W}=e$.

In the $\mu\nu{\rm SSM}$, the masses of left-handed sneutrinos are basically determined by $m_{\tilde L_i}$, and the three right-handed sneutrinos are essentially degenerated. The CP-even and CP-odd right-handed sneutrinos mass squared $m_{S_{5+i}}^2$ and $m_{P_{5+i}}^2$ could be approximately written as
\begin{eqnarray}
&&m_{S_{5+i}}^2\approx (A_\kappa+4\kappa\upsilon_{\nu^c})\kappa\upsilon_{\nu^c} +A_\lambda \lambda \upsilon_d \upsilon_u/\upsilon_{\nu^c}-2\lambda^2(\upsilon_d^2+\upsilon_u^2),\nonumber\\
&&m_{P_{5+i}}^2\approx -3A_\kappa \kappa\upsilon_{\nu^c} +(A_\lambda/\upsilon_{\nu^c}+4\kappa)\lambda \upsilon_d \upsilon_u-2\lambda^2(\upsilon_d^2+\upsilon_u^2).
\end{eqnarray}
Here, the main contribution to the mass squared is the first term as $\kappa$ is large, due to $\upsilon_{\nu^c} \gg \upsilon_{u,d}$. Therefore, we could use the approximate relation
\begin{eqnarray}
-4\kappa\upsilon_{\nu^c}\lesssim A_\kappa \lesssim 0
\label{tachyon}
\end{eqnarray}
to avoid the tachyons.

Before the numerical calculation, the constraints on the parameters of the $\mu\nu{\rm SSM}$ from neutrino experiments should be considered at first. Three flavor neutrinos $\nu_{e,\mu,\tau}$ could mix into three massive neutrinos $\nu_{1,2,3}$ during their flight, and the mixing is described by the Pontecorvo-Maki-Nakagawa-Sakata unitary matrix $U_{_{PMNS}}$ \cite{neutrino-oscillations}. Through several recent reactor oscillation experiments~\cite{theta13}, $\theta_{13}$ is now precisely known. The global fit of $\theta_{13}$ gives~\cite{Garcia}
\begin{eqnarray}
\sin^2\theta_{13}=0.023\pm 0.0023.
\label{neutrino-oscillations1}
\end{eqnarray}
The other experimental observations of the parameters in $U_{_{PMNS}}$ for the normal mass hierarchy~\cite{Garcia} show that
\begin{eqnarray}
&&\sin^2\theta_{12} =0.302_{-0.012}^{+0.013},\qquad  \Delta m_{21}^2 =7.50_{-0.19}^{+0.18}\times 10^{-5} {\rm eV}^2,  \nonumber\\
&&\sin^2\theta_{23}=0.413_{-0.025}^{+0.037},\qquad  \Delta m_{31}^2 =2.473_{-0.067}^{+0.070}\times 10^{-3} {\rm eV}^2.
\label{neutrino-oscillations2}
\end{eqnarray}

In the $\mu\nu{\rm SSM}$, the three neutrino masses are obtained through a TeV scale seesaw mechanism \cite{mnSSM,neutrino-mass}. Assuming that the charged lepton mass matrix in the flavor basis is in the diagonal form, we parametrize the unitary matrix which diagonalizes the effective neutrino mass matrix $m_{eff}$ (see Ref.~\cite{Zhang}) as \cite{Uv}
\begin{eqnarray}
{U_\nu} = &&\left( {\begin{array}{*{20}{c}}
   {{c_{12}}{c_{13}}} & {{s_{12}}{c_{13}}} & {{s_{13}}{e^{ - i\delta }}}  \\
   { - {s_{12}}{c_{23}} - {c_{12}}{s_{23}}{s_{13}}{e^{i\delta }}} & {{c_{12}}{c_{23}} - {s_{12}}
   {s_{23}}{s_{13}}{e^{i\delta }}} & {{s_{23}}{c_{13}}}  \\
   {{s_{12}}{s_{23}} - {c_{12}}{c_{23}}{s_{13}}{e^{i\delta }}} & { - {c_{12}}{s_{23}} - {s_{12}}
   {c_{23}}{s_{13}}{e^{i\delta }}} & {{c_{23}}{c_{13}}}  \\
\end{array}} \right)  \nonumber\\
&&\times \: diag(1,{e^{i\frac{{{\alpha _{21}}}}{2}}},{e^{i\frac{{{\alpha _{31}}}}{2}}})\:,
\end{eqnarray}
where ${c_{_{ij}}} = \cos {\theta _{ij}}$, ${s_{_{ij}}} = \sin {\theta _{ij}}$. In the next calculation, the values of $\theta_{ij}$ are obtained from the experimental data in Eq.~(\ref{neutrino-oscillations1}) and Eq.~(\ref{neutrino-oscillations2}), and all CP violating phases $\delta$, $\alpha _{21}$, and $\alpha _{31}$ are set to zero. The unitary matrix  $U_\nu$ diagonalizes the effective neutrino mass matrix $m_{eff}$ in the following way:
\begin{eqnarray}
U_\nu ^T m_{eff}^T{m_{eff}}{U_\nu} = diag({m_{\nu _1}^2},{m_{\nu _2}^2},{m_{\nu _3}^2}).
\label{eff}
\end{eqnarray}
For the neutrino mass spectrum, we assume it to be normal hierarchical, i.e., ${m_{\nu_1}}{\rm{ < }}{m_{\nu_2}}{\rm{ < }}{m_{\nu_3}}$, and we choose the lightest neutrino mass $m_{\nu_1}=10^{-2}\:{\rm{eV}}$ as input in our numerical analysis, limited by neutrino masses from neutrinoless double-$\beta$ decay~\cite{neu-m-limit} and cosmology~\cite{neu-m-limit1}. The other two neutrino masses $m_{\nu_{2,3}}$ can be obtained through the experimental data on the differences of neutrino mass squared in Eq.~(\ref{neutrino-oscillations2}).  Then we can numerically derive $Y_{\nu_i} \sim \mathcal{O}(10^{-7})$ and $a_{\nu_i} \sim \mathcal{O}(-10^{-4}{\rm{GeV}})$ from Eq.~(\ref{eff}). Accordingly, $\upsilon_{\nu_i} \sim \mathcal{O}(10^{-4}{\rm{GeV}})$ through Eq.~(\ref{eq-min}). Due to $\upsilon_{\nu_i}\ll\upsilon_{u,d}$, we can have
\begin{eqnarray}
\tan\beta\simeq \frac{\upsilon_u}{\upsilon_d}.
\end{eqnarray}

We also impose a constraint on the SUSY contribution to the muon anomalous magnetic dipole moment $a_\mu$ in the $\mu\nu$SSM~\cite{Zhang}. The difference between experiment and the SM prediction on $a_\mu$ is~\cite{PDG,E821}
\begin{eqnarray}
\Delta a_\mu =a_\mu^{{\rm{exp}}} -a_\mu^{{\rm{SM}}} = (24.8\pm7.9)\times 10^{-10},
\end{eqnarray}
with all errors combined in quadrature. Therefore, the SUSY contribution to $a_\mu$ in the $\mu\nu$SSM should be constrained as $1.1\times 10^{-10} \leq \Delta a_\mu \leq 48.5\times 10^{-10}$, where a $3 \sigma$ experimental error is considered. In Ref.~\cite{Zhang}, we can know that the experimental data for $a_\mu$ will give a large constraint on the parameter $M_2$, for a given value of $\tan \beta$.

For relevant parameters in the SM, we choose~\cite{PDG}
\begin{eqnarray}
&&\alpha_s(m_{Z})=0.118,\qquad  m_t=173.5\;{\rm GeV},\qquad  m_{Z}=91.188\;{\rm GeV},
\nonumber\\
&&\alpha(m_{Z})=1/128,\qquad\,  m_b=4.65\;{\rm GeV}, \qquad\: m_{W}=80.385\;{\rm GeV}.
\end{eqnarray}
Through the analysis of the parameter space in Ref.~\cite{mnSSM1}, we could choose the reasonable values for some parameters in the $\mu\nu{\rm SSM}$ as $\kappa=0.4$, ${A_{\kappa}}=-300\;{\rm GeV}$, and $A_{\lambda}=500\;{\rm GeV}$ for simplicity. Here we choose small ${A_{\kappa}}$ to avoid the tachyons, through Eq.~(\ref{tachyon}). We assume that the first two generations of squarks and the right-handed sbottom are heavy, $m_{{\tilde Q}_{1,2}}=m_{{\tilde u}^c_{1,2}}=m_{{\tilde d}^c_{1,2,3}}=2\;{\rm TeV}$, because they play a minor role for the Higgs physics. For simplicity, we can choose $m_{{\tilde L}_{1,2}}=m_{{\tilde e}^c_{1,2}}=1\;{\rm TeV}$ and $A_{e_{1,2,3}}=A_{d_{1,2,3}}=A_{u_{1,2}}=1\;{\rm TeV}$. As key parameters, $m_{{\tilde Q}_3}$, $m_{{\tilde u}^c_3}$ and $A_{u_3}\equiv A_t$, affects the 125 GeV Higgs mass and decays.

Stops have been searched for at the LHC in gauge-mediated SUSY breaking (GMSB) models~\cite{stop-GMSB}, where the gravitino ($\tilde{G}$) is typically the LSP which is similar to the $\mu\nu{\rm SSM}$. These studies rule out stop masses up to 200--600 GeV, where the light stop $\tilde t_1$ might decay via $b\tilde{\chi}_1^\pm$, $t\tilde{\chi}_1^0$ and $\tilde{\chi}_1^0$ decay in $Z(h)\tilde{G}$. These studies assumed that the lightest neutralino mass is less than the light stop mass, $m_{\tilde{\chi}_1^0}<m_{{\tilde t}_1}$. The  $m_{\tilde{\chi}_1^0}>m_{{\tilde t}_1}$ case still needs be tested in the future. So, for $m_{\tilde{\chi}_1^0}>m_{{\tilde t}_1}$, we still could consider $m_{{\tilde t}_1}<600$ GeV to study the light stop effect on Higgs decays in the $\mu\nu{\rm SSM}$. Constrained by the 125 GeV Higgs, we could have a several TeV heavy stop and a several hundred GeV light stop. To keep the left-handed sbottom heavy, we choose $m_{{\tilde Q}_3}\gg m_{{\tilde u}_3^c}$~\cite{Higgs-Carena1}. In the following, we take $m_{{\tilde Q}_3}=2\;{\rm TeV}$ for simplicity. So, here the heavy stop mass $m_{{\tilde t}_2}$ is around 2 TeV.
Then, the free parameters that affect our next analysis are
\begin{eqnarray}
\tan \beta ,\;  \lambda, \; {\upsilon_{\nu^c}},\; M_2,\; {A_{t}},\; m_{{\tilde u}_3^c},\; m_{{\tilde L}_3} ,\; m_{{\tilde e}_3^c} .
\end{eqnarray}

\begin{table*}
\begin{tabular*}{\textwidth}{@{\extracolsep{\fill}}lllll@{}}
\hline
Parameters&Min&Max&Step\\
\hline
$\tan \beta$&2&30&7\\
$ \lambda$&0.1&0.2&0.05\\
${\upsilon_{\nu^c}}/{\rm TeV}$&1&3&1\\
$M_2/{\rm TeV}$&0.5&3.5&1\\
$A_{t}/{\rm TeV}$&-2.6&3.4&0.1\\
$m_{{\tilde u}_3^c}/{\rm GeV}$&100&800&20\\
\hline
\end{tabular*}
\caption{Scanning parameters for the light stop effect on the Higgs decays.}
\label{tab2}
\end{table*}

\begin{figure}
\setlength{\unitlength}{1mm}
\centering
\begin{minipage}[c]{0.45\textwidth}
\includegraphics[width=2.6in]{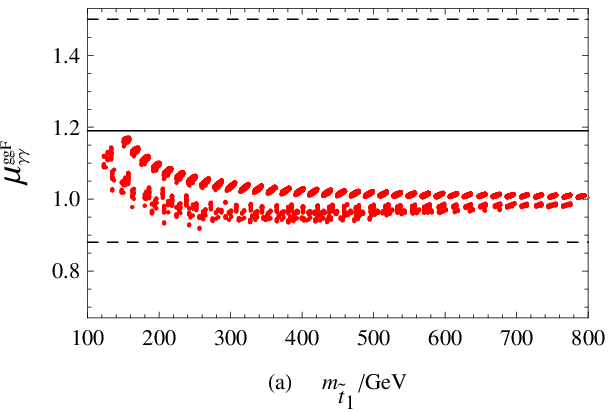}
\end{minipage}%
\begin{minipage}[c]{0.45\textwidth}
\includegraphics[width=2.6in]{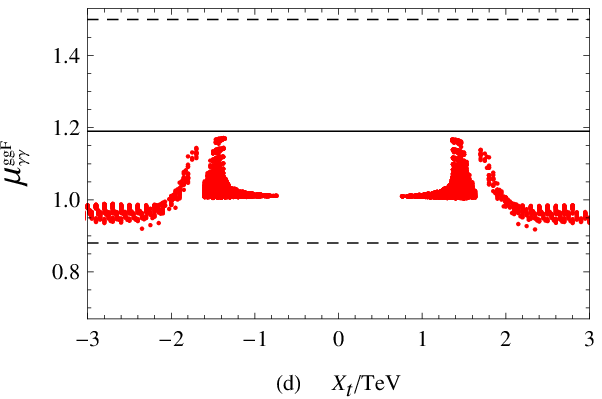}
\end{minipage}
\begin{minipage}[c]{0.45\textwidth}
\includegraphics[width=2.6in]{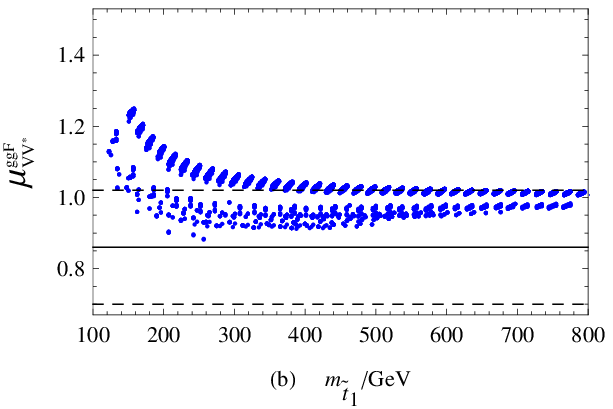}
\end{minipage}%
\begin{minipage}[c]{0.45\textwidth}
\includegraphics[width=2.6in]{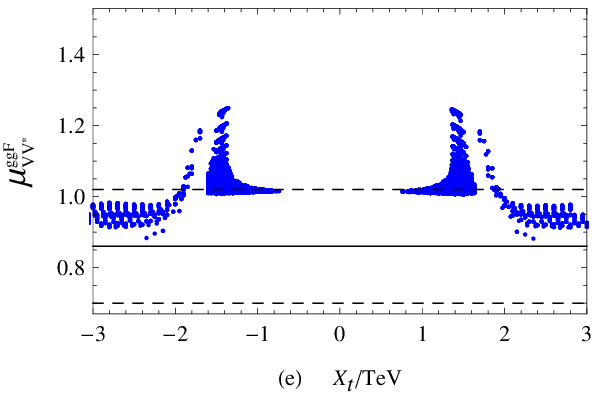}
\end{minipage}
\begin{minipage}[c]{0.45\textwidth}
\includegraphics[width=2.6in]{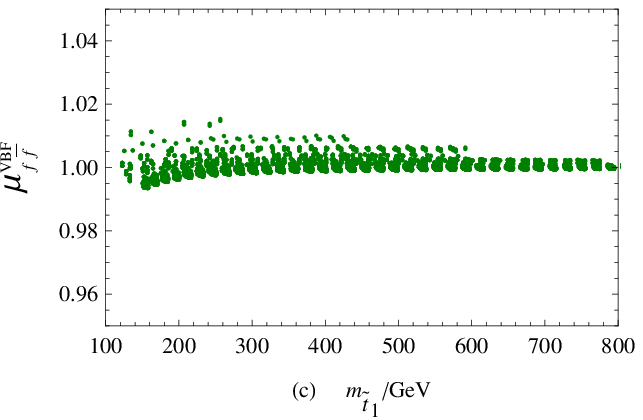}
\end{minipage}%
\begin{minipage}[c]{0.45\textwidth}
\includegraphics[width=2.6in]{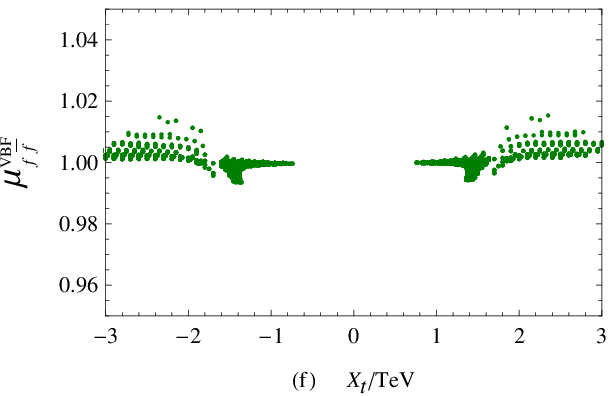}
\end{minipage}
\caption[]{(Color online) $\mu_{\gamma\gamma}^{{\rm{ggF}}}$ (a),(d), $\mu_{VV^*}^{{\rm{ggF}}}$ (b),(e) and $\mu_{f\bar{f}}^{{\rm{VBF}}}$ (c),(f) vary with $m_{{\tilde t}_1}$ and $X_t= A_t-\mu/\tan\beta$, respectively, where the horizontal solid lines correspond to the experimental central values and the dashed lines to the $1\sigma$ intervals.}
\label{fig-st}
\end{figure}

Taking $m_{{\tilde L}_3}= m_{{\tilde e}_3^c}=1\;{\rm TeV}$ to ignore the light stau effect, we study the light stop effect on Higgs decays in the $\mu\nu{\rm SSM}$ in Fig.~\ref{fig-st}, by scanning the parameters listed in Table~\ref{tab2}. In Table~\ref{tab2}, we take relatively small value of the parameter $\lambda$, considering the Landau pole condition at the high-energy scale~\cite{mnSSM1}. In the scanning, we avoid the tachyons, simultaneously coinciding with $m_{\tilde{\chi}_1^0}>m_{{\tilde t}_1}$, and the heavy doubletlike Higgs mass $m_H\geq 642$ GeV~\cite{mH-642}. The results are also constrained by the light doubletlike Higgs mass with $124.5\,{\rm GeV}\leq m_{{h}} \leq126.9\:{\rm GeV}$ and the muon anomalous magnetic dipole moment $1.1\times 10^{-10} \leq \Delta a_\mu \leq 48.5\times 10^{-10}$, where a $3 \sigma$ experimental error is considered.

In Fig.~\ref{fig-st}, we plot the signal strengths $\mu_{\gamma\gamma}^{{\rm{ggF}}}$ (a), $\mu_{VV^*}^{{\rm{ggF}}}$ (b), and $\mu_{f\bar{f}}^{{\rm{VBF}}}$ (c) varying with the light stop mass $m_{{\tilde t}_1}$, respectively, where the horizontal solid lines correspond to the experimental central values and the dashed lines to the $1\sigma$ intervals. The numerical results show that the light stop could give large effect on the signal strengths $\mu_{\gamma\gamma}^{{\rm{ggF}}}$ and $\mu_{VV^*}^{{\rm{ggF}}}$, as the light stop mass is small.  With increasing of the light stop mass, the contribution of the light stop for the signal strengths become small. When $m_{{\tilde t}_1}\gtrsim 700\;{\rm GeV}$, the signal strengths $\mu_{\gamma\gamma}^{{\rm{ggF}}}$ and $\mu_{VV^*}^{{\rm{ggF}}}$ are close to 1, which is in agreement with the SM. Fig.~\ref{fig-st}(c) indicates the light stop play a minor role for the signal strength $\mu_{f\bar{f}}^{{\rm{VBF}}}$.

To explain the results of the signal strengths further, in Fig.~\ref{fig-st} we also plot the signal strengths $\mu_{\gamma\gamma}^{{\rm{ggF}}}$ (d), $\mu_{VV^*}^{{\rm{ggF}}}$ (e), and $\mu_{f\bar{f}}^{{\rm{VBF}}}$ (f), respectively, versus $X_t= A_t-\mu/\tan\beta$, where $\mu=3\lambda{\upsilon_{\nu^c}}$. One can find that the signal strengths $\mu_{\gamma\gamma,VV^*}^{{\rm{ggF}}}>1$ when $|X_t|\lesssim 2$ TeV and  $\mu_{\gamma\gamma,VV^*}^{{\rm{ggF}}}<1$ for $|X_t|\gtrsim 2$ TeV. This shows that the light stop effect can be of either sign, depending on the parameter $X_t= A_t-\mu/\tan\beta$, as we will discuss in detail below. Coinciding with the MSSM, the stop loop contributions to the $gg$ or $\gamma\gamma$ amplitude  in the $\mu\nu{\rm SSM}$ can be approximately proportional to~\cite{Higgs-Arvanitaki,Higgs-Blum,Higgs-Buckley,Higgs-Espinosa,Higgs-Carena2}
\begin{eqnarray}
\Delta \mathcal{A}_{gg,\gamma\gamma}^{{\tilde t}}\propto
\frac{m_t^2}{m_{{\tilde t}_1}^2 m_{{\tilde t}_2}^2}(m_{{\tilde t}_1}^2+m_{{\tilde t}_2}^2-X_t^2),
\end{eqnarray}
For $X_t^2<(m_{{\tilde t}_1}^2+m_{{\tilde t}_2}^2)$, the stops lead to an enhancement of the gluon-gluon Higgs production. So, the signal strengths $\mu_{\gamma\gamma,VV^*}^{{\rm{ggF}}}>1$, when $|X_t|<\sqrt{(m_{{\tilde t}_1}^2+m_{{\tilde t}_2}^2)}\sim2$ TeV. In Fig.~\ref{fig-st}, the signal strength $\mu_{VV^*}^{{\rm{ggF}}}$ can reach 1.25; however, the signal strength $\mu_{\gamma\gamma}^{{\rm{ggF}}}$ just reaches 1.17, since the stops lead to a reduction of the Higgs to diphoton decay width for $X_t^2<(m_{{\tilde t}_1}^2+m_{{\tilde t}_2}^2)$. On the contrary, the stops reduce the signal strengths $\mu_{\gamma\gamma}^{{\rm{ggF}}}$ and  $\mu_{VV^*}^{{\rm{ggF}}}$, when $X_t^2>(m_{{\tilde t}_1}^2+m_{{\tilde t}_2}^2)$. Additionally, the signal strength $\mu_{f\bar{f}}^{{\rm{VBF}}}<1$ as $|X_t|\lesssim 2$ TeV and $\mu_{f\bar{f}}^{{\rm{VBF}}}>1$ when $|X_t|\gtrsim 2$ TeV, is due to be rescaled by the total width ${\Gamma_{{\rm{SM}}}^h / \Gamma_{{\rm{NP}}}^h}$ in Eq.~(\ref{signals}).

\begin{table*}
\begin{tabular*}{\textwidth}{@{\extracolsep{\fill}}lllll@{}}
\hline
Parameters&Min&Max&Step\\
\hline
$\tan \beta$&30&60&5\\
$ \lambda$&0.1&0.2&0.05\\
${\upsilon_{\nu^c}}/{\rm TeV}$&1&3&0.2\\
$M_2/{\rm TeV}$&0.5&3.5&1.5\\
$A_{t}/{\rm TeV}$&-2.6&3.4&0.2\\
$m_{{\tilde e}_3^c}/{\rm GeV}$&100&400&30\\
\hline
\end{tabular*}
\caption{Scanning parameters for the light stau effect on the Higgs to diphoton decay.}
\label{tab3}
\end{table*}

\begin{figure}
\setlength{\unitlength}{1mm}
\centering
\begin{minipage}[c]{0.45\textwidth}
\includegraphics[width=2.6in]{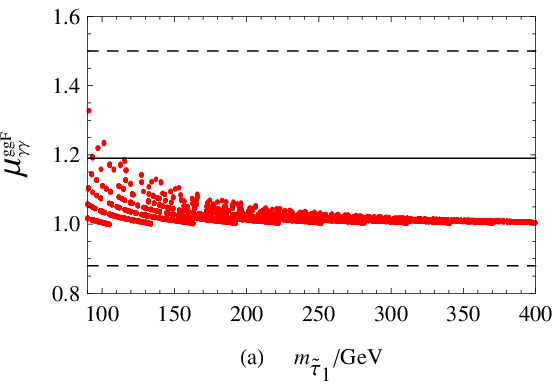}
\end{minipage}%
\begin{minipage}[c]{0.45\textwidth}
\includegraphics[width=2.6in]{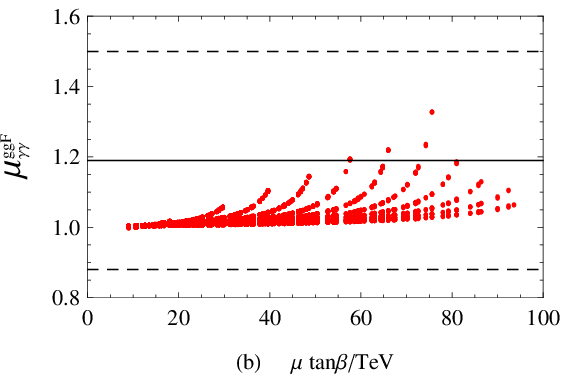}
\end{minipage}
\caption[]{(Color online) $\mu_{\gamma\gamma}^{{\rm{ggF}}}$ vary with $m_{{\tilde \tau}_1}$ (a) and  $\mu \tan \beta$ (b), respectively, where the horizontal solid lines correspond to the experimental central values given in Table~\ref{tab1} and the dashed lines to the $1\sigma$ intervals.}
\label{fig-stau}
\end{figure}

Taking $m_{{\tilde u}_3^c}=1\;{\rm TeV}$ to ignore the stop effect and choosing $m_{{\tilde L}_3}=0.5\;{\rm TeV}$ to keep the third generation of left-handed sneutrinos relatively heavy, we study the light stau effect on the Higgs to diphoton decay in the $\mu\nu{\rm SSM}$ in Fig.~\ref{fig-stau}, where we scan the parameter space listed in Table~\ref{tab3}. In the scanning, the results are also constrained by the heavy doubletlike Higgs mass $m_H\geq 642$ GeV, the light doubletlike Higgs mass with $124.5\,{\rm GeV}\leq m_{{h}} \leq126.9\:{\rm GeV}$, and the muon anomalous magnetic dipole moment. Here we consider the constraint of the light stau mass $m_{{\tilde \tau}_1}\gtrsim 90$ GeV from the LEP limit. In Fig.~\ref{fig-stau}, we plot the signal strength $\mu_{\gamma\gamma}^{{\rm{ggF}}}$ varying with the light stau mass $m_{{\tilde \tau}_1}$ (a) and $\mu \tan \beta$ (b), respectively. Figure~\ref{fig-stau}(a) shows that the light stau can give a large enhancement on the signal strength $\mu_{\gamma\gamma}^{{\rm{ggF}}}$, when $m_{{\tilde \tau}_1}\lesssim300\;{\rm GeV}$. Figure~\ref{fig-stau}(b) indicates that the signal strength $\mu_{\gamma\gamma}^{{\rm{ggF}}}$ is enhanced greatly, as $\mu \tan \beta$ is large because large values of $\mu$ and $\tan \beta$ induce large mixing in the stau sector leading to an enhancement of the Higgs to diphoton decay width~\cite{Higgs-Blum,Higgs-Carena2,Higgs-Carena3}.

\section{Summary\label{sec5}}

In the framework of the $\mu\nu$SSM, we attempt to account for the experimental data on the Higgs reported by ATLAS and CMS recently. Under some assumptions and constraints of the parameter space, the results indicate that the 125 GeV Higgs decay signal strengths $\mu_{\gamma\gamma}^{{\rm{ggF}}}$, $\mu_{VV^\ast}^{{\rm{ggF}}}$  ($V=Z,W$) and $\mu_{f\bar{f}}^{{\rm{VBF}}}$ $(f=b,\tau)$ can fit the experimental data. Meanwhile, the numerical evaluations on the heavy doubletlike Higgs  mass $m_H$ exceed 642 GeV.

In the $\mu\nu$SSM, we show the light stop and stau contributions to the 125 GeV Higgs decay signal strengths. The light stop leads to an enhancement or  reduction  of the signal strengths $\mu_{\gamma\gamma}^{{\rm{ggF}}}$ and $\mu_{VV^*}^{{\rm{ggF}}}$. The signal strength $\mu_{f\bar{f}}^{{\rm{VBF}}}$ is consistent with the SM. For large $\mu$ and $\tan \beta$, the light stau could considerably enhance the signal strength $\mu_{\gamma\gamma}^{{\rm{ggF}}}$. Note that for the signal strengths $\mu_{\gamma\gamma,ZZ^*}^{{\rm{ggF}}}$, ATLAS and CMS report currently quite different central values, as indicated in Table~\ref{tab1}. Here the average of the two values is just used as a rough guideline. In the near future, further constraints can be obtained from more precise determinations of the signal strengths in the measured decay channels at the LHC.

\section*{Acknowledgements}
This work has been supported by the National Natural Science Foundation of China (NNSFC)
with Grants No. 11275036 and No. 11047002, the open project of State
Key Laboratory of Mathematics-Mechanization with Grant No. Y3KF311CJ1, the Natural
Science Foundation of Hebei province with Grant No. A2013201277, and the Natural Science Fund of Hebei University with Grants No. 2011JQ05 and No. 2012-242.

\appendix

\section{The radiative corrections\label{app-rad}}

The radiative corrections on the doubletlike Higgses originate from fermions and corresponding supersymmetric partners in the $\mu\nu$SSM:
\begin{eqnarray}
&&\Delta_{11}=\Delta_{11}^{t}+\Delta_{11}^{b}+\Delta_{11}^{l},
\nonumber\\
&&\Delta_{12}=\Delta_{12}^{t}+\Delta_{11}^{b}+\Delta_{12}^{l},\nonumber\\
&&\Delta_{22}=\Delta_{22}^{t}+\Delta_{11}^{b}+\Delta_{22}^{l}.
\end{eqnarray}
Neglecting the terms containing small coupling $Y_{\nu_i}$ and $\upsilon_{\nu_i}$, and using the expressions given in Ref.~\cite{Higgs-loop}, the radiative corrections from the top quark and its scalar partner $\tilde{t}_{1,2}$ including two-loop leading-log effects~\cite{Higgs-Carena} read as
\begin{eqnarray}
&&\Delta_{11}^{t}=\frac{3G_F{m_{t}^4}}{2\sqrt{2}\pi^2\sin^2\beta}{\mu^2(A_{t}-\mu\cot\beta)^2\over {(m_{\tilde{t}_1}^2-m_{\tilde{t}_2}^2)}^2}
g(m_{\tilde{t}_1}^2,m_{\tilde{t}_2}^2),
\nonumber\\
&&\Delta_{12}^{t}=\frac{3G_F{m_{t}^4}}{2\sqrt{2}\pi^2\sin^2\beta}{\mu(-A_{t}+\mu\cot\beta)\over m_{\tilde{t}_1}^2-m_{\tilde{t}_2}^2}\Big\{ \ln{m_{\tilde{t}_1}^2\over m_{\tilde{t}_2}^2}
+{A_{t}(A_{t}-\mu\cot\beta)\over {(m_{\tilde{t}_1}^2-m_{\tilde{t}_2}^2)}}
g(m_{\tilde{t}_1}^2,m_{\tilde{t}_2}^2)\Big\},
\nonumber\\
&&\Delta_{22}^{t}=\frac{3G_F{m_{t}^4}}{2\sqrt{2}\pi^2\sin^2\beta}\Big\{\ln{m_{\tilde{t}_1}^2 m_{\tilde{t}_2}^2 \over m_{t}^4}+{2A_{t}(A_{t}-\mu\cot\beta)\over m_{\tilde{t}_1}^2-m_{\tilde{t}_2}^2} \ln{m_{\tilde{t}_1}^2\over m_{\tilde{t}_2}^2}
\nonumber\\
&&\hspace{1.3cm}
+{A_{t}^2(A_{t}-\mu\cot\beta)^2\over {(m_{\tilde{t}_1}^2-m_{\tilde{t}_2}^2)}^2}
g(m_{\tilde{t}_1}^2,m_{\tilde{t}_2}^2)+ {1\over {16 \pi^2}}\ln{m_{\tilde{t}_1}^2 m_{\tilde{t}_2}^2 \over m_{t}^4} \Big(\frac{3e^2m_t^2}{4s_W^2 m_W^2}-32\pi \alpha_s \Big) \nonumber\\
&&\hspace{1.3cm}
\times \Big[{1\over {2}}\ln{m_{\tilde{t}_1}^2 m_{\tilde{t}_2}^2 \over m_{t}^4} + {2(A_{t}-\mu\cot\beta)^2\over {m_{\tilde{t}_1}m_{\tilde{t}_2}}} \Big( 1-{(A_{t}-\mu\cot\beta)^2\over {12 m_{\tilde{t}_1}m_{\tilde{t}_2}}} \Big) \Big] \Big\},
\end{eqnarray}
with
\begin{eqnarray}
g(m_1^2,m_2^2)=2-{m_1^2+m_2^2\over m_1^2-m_2^2}\ln{m_1^2\over m_2^2}.
\end{eqnarray}
The one-loop radiative corrections from the bottom quark and its scalar partner $\tilde{b}_{1,2}$ are formulated as
\begin{eqnarray}
&&\Delta_{11}^{b}=\frac{3G_F{m_{b}^4}}{2\sqrt{2}\pi^2\cos^2\beta}\Big\{\ln{m_{\tilde{b}_1}^2 m_{\tilde{b}_2}^2 \over m_{b}^4}+{2A_{b}(A_{b}-\mu\tan\beta)\over m_{\tilde{b}_1}^2-m_{\tilde{b}_2}^2} \ln{m_{\tilde{b}_1}^2\over m_{\tilde{b}_2}^2}
\nonumber\\
&&\hspace{1.3cm}
+{A_{b}^2(A_{b}-\mu\tan\beta)^2\over {(m_{\tilde{b}_1}^2-m_{\tilde{b}_2}^2)}^2}
g(m_{\tilde{b}_1}^2,m_{\tilde{b}_2}^2)\Big\},
\nonumber\\
&&\Delta_{12}^{b}=\frac{3G_F{m_{b}^4}}{2\sqrt{2}\pi^2\cos^2\beta}{\mu(-A_{b}+\mu\tan\beta)\over m_{\tilde{b}_1}^2-m_{\tilde{b}_2}^2} \Big\{\ln{m_{\tilde{b}_1}^2\over m_{\tilde{b}_2}^2} +{A_{b}(A_{b}-\mu\tan\beta)\over {(m_{\tilde{b}_1}^2-m_{\tilde{b}_2}^2)}}
g(m_{\tilde{b}_1}^2,m_{\tilde{b}_2}^2)\Big\},
\nonumber\\
&&\Delta_{22}^{b}=\frac{3G_F{m_{b}^4}}{2\sqrt{2}\pi^2\cos^2\beta}{\mu^2(A_{b}-\mu\tan\beta)^2\over {(m_{\tilde{b}_1}^2-m_{\tilde{b}_2}^2)}^2}g(m_{\tilde{b}_1}^2,m_{\tilde{b}_2}^2),
\end{eqnarray}
Similarly, one can obtain the one-loop radiative corrections from the $\tau$ lepton and its scalar partner $\tilde{\tau}_{1,2}$:
\begin{eqnarray}
&&\Delta_{11}^{l}=\frac{G_F{m_{\tau}^4}}{2\sqrt{2}\pi^2\cos^2\beta}\Big\{\ln{m_{\tilde{\tau}_1}^2 m_{\tilde{\tau}_2}^2 \over m_{\tau}^4}+{2A_{\tau}(A_{\tau}-\mu\tan\beta)\over m_{\tilde{\tau}_1}^2-m_{\tilde{\tau}_2}^2} \ln{m_{\tilde{\tau}_1}^2\over m_{\tilde{\tau}_2}^2}
\nonumber\\
&&\hspace{1.3cm}
+{A_{\tau}^2(A_{\tau}-\mu\tan\beta)^2\over {(m_{\tilde{\tau}_1}^2-m_{\tilde{\tau}_2}^2)}^2}
g(m_{\tilde{\tau}_1}^2,m_{\tilde{\tau}_2}^2)\Big\},
\nonumber\\
&&\Delta_{12}^{l}=\frac{G_F{m_{\tau}^4}}{2\sqrt{2}\pi^2\cos^2\beta}{\mu(-A_{\tau}+\mu\tan\beta)\over m_{\tilde{\tau}_1}^2-m_{\tilde{\tau}_2}^2} \Big\{\ln{m_{\tilde{\tau}_1}^2\over m_{\tilde{\tau}_2}^2} +{A_{\tau}(A_{\tau}-\mu\tan\beta)\over {(m_{\tilde{\tau}_1}^2-m_{\tilde{\tau}_2}^2)}}
g(m_{\tilde{\tau}_1}^2,m_{\tilde{\tau}_2}^2)\Big\},
\nonumber\\
&&\Delta_{22}^{l}=\frac{G_F{m_{\tau}^4}}{2\sqrt{2}\pi^2\cos^2\beta}{\mu^2(A_{\tau}-\mu\tan\beta)^2\over {(m_{\tilde{\tau}_1}^2-m_{\tilde{\tau}_2}^2)}^2}
g(m_{\tilde{\tau}_1}^2,m_{\tilde{\tau}_2}^2).
\end{eqnarray}

\section{The couplings\label{app-coupling}}

The couplings of CP-even neutral scalars and charged scalars are formulated as
\begin{eqnarray}
\mathcal{L}_{int} = \sum\limits_{\alpha,\beta,\gamma=1}^8 C_{\alpha \beta \gamma}^{S^{\pm}} S_\alpha S_\beta^+ S_\gamma^-,
\end{eqnarray}
with
\begin{eqnarray}
&&C_{\alpha \beta \gamma}^{S^{\pm}} = \frac{-e^2}{2\sqrt{2}{s_{_W}^2}} \Big[ \upsilon_d R_S^{1\alpha}R_{S^\pm}^{1\beta}R_{S^\pm}^{1\gamma} +(\upsilon_d R_S^{2\alpha} + \upsilon_u R_S^{1\alpha}) R_{S^\pm}^{1\beta}R_{S^\pm}^{2\gamma}   \nonumber\\
&&\hspace{1.4cm}
+ \upsilon_u R_S^{2\alpha}R_{S^\pm}^{2\beta}R_{S^\pm}^{2\gamma} +(\upsilon_d R_S^{(2+i)\alpha} + \upsilon_{\nu_i} R_S^{1\alpha}) R_{S^\pm}^{(2+i)\beta}R_{S^\pm}^{1\gamma}  \nonumber\\
&&\hspace{1.4cm}
+ (\upsilon_u R_S^{(2+i)\alpha} + \upsilon_{\nu_i} R_S^{2\alpha}) R_{S^\pm}^{(2+i)\beta}R_{S^\pm}^{2\gamma} + \upsilon_{\nu_i} R_S^{(2+j)\alpha}R_{S^\pm}^{(2+i)\beta}R_{S^\pm}^{(2+j)\gamma} \Big]  \nonumber\\
&&\hspace{1.4cm}
+  \frac{e^2}{4\sqrt{2}{s_{_W}^2}c_{_W}^2} ( \upsilon_d R_S^{1\alpha} - \upsilon_u R_S^{2\alpha} + \upsilon_{\nu_i} R_S^{(2+i)\alpha} ) \Big[(c_{_W}^2-s_{_W}^2) \delta^{\beta\gamma}   \nonumber\\
&&\hspace{1.4cm} -2(c_{_W}^2-s_{_W}^2) R_{S^\pm}^{2\beta}R_{S^\pm}^{2\gamma} - (c_{_W}^2-3s_{_W}^2) R_{S^\pm}^{(5+j)\beta}R_{S^\pm}^{(5+j)\gamma} \Big]   \nonumber\\
&&\hspace{1.4cm}
-\frac{1}{\sqrt{2}}\lambda_i \lambda_j\upsilon_{\nu_i}R_S^{(5+j)\alpha} \Big[ R_{S^\pm}^{1\beta}R_{S^\pm}^{1\gamma} + R_{S^\pm}^{2\beta}R_{S^\pm}^{2\gamma} \Big]  \nonumber\\
&&\hspace{1.4cm}
+\frac{1}{\sqrt{2}}\lambda_i {Y_{\nu_{kj}}}(\upsilon_{\nu_i^c} R_S^{(5+j)\alpha}+\upsilon_{\nu_j^c} R_S^{(5+i)\alpha}) R_{S^\pm}^{(2+k)\beta}R_{S^\pm}^{1\gamma}\nonumber\\
&&\hspace{1.4cm}
+\frac{1}{\sqrt{2}}\lambda_k {Y_{e_{ij}}}  (\upsilon_{\nu_i} R_S^{(5+k)\alpha}+\upsilon_{\nu_k^c} R_S^{(2+i)\alpha}) R_{S^\pm}^{2\beta}R_{S^\pm}^{(5+j)\gamma} \nonumber\\
&&\hspace{1.4cm}+\frac{1}{\sqrt{2}}\lambda_k {Y_{e_{ij}}}(\upsilon_u R_S^{(5+k)\alpha}+\upsilon_{\nu_k^c} R_S^{2\alpha}) R_{S^\pm}^{(2+i)\beta} R_{S^\pm}^{(5+j)\gamma} \nonumber\\
&&\hspace{1.4cm}
-\frac{1}{\sqrt{2}}{Y_{\nu_{ki}}}\upsilon_{\nu_i^c} R_S^{(5+j)\alpha} \Big[ {Y_{\nu_{lj}}} R_{S^\pm}^{(2+l)\beta}R_{S^\pm}^{(2+k)\gamma}+ {Y_{\nu_{kj}}} R_{S^\pm}^{2\beta}R_{S^\pm}^{2\gamma} \Big]  \nonumber\\
&&\hspace{1.4cm}
+\frac{1}{\sqrt{2}}{Y_{e_{ki}}} {Y_{e_{kj}}} \Big[ \upsilon_d R_S^{1\alpha} (R_{S^\pm}^{(2+i)\beta}R_{S^\pm}^{(2+j)\gamma}-R_{S^\pm}^{(5+i)\beta}R_{S^\pm}^{(5+j)\gamma}) \nonumber\\
&&\hspace{1.4cm}
-(\upsilon_d R_S^{(2+i)\alpha} + \upsilon_{\nu_i} R_S^{1\alpha}) R_{S^\pm}^{(2+j)\beta}R_{S^\pm}^{1\gamma} + \upsilon_{\nu_i} R_S^{(2+j)\alpha} R_{S^\pm}^{1\beta}R_{S^\pm}^{1\gamma} \Big] \nonumber\\
&&\hspace{1.4cm}
+\frac{1}{\sqrt{2}}{Y_{e_{ki}}} {Y_{\nu_{kj}}} (\upsilon_d R_S^{(5+i)\alpha}+\upsilon_{\nu_i^c} R_S^{1\alpha}) R_{S^\pm}^{2\beta} R_{S^\pm}^{(5+j)\gamma} \nonumber\\
&&\hspace{1.4cm}
+\frac{1}{\sqrt{2}}{Y_{e_{ki}}} {Y_{\nu_{kj}}} (\upsilon_u R_S^{(5+i)\alpha}+\upsilon_{\nu_i^c} R_S^{2\alpha}) R_{S^\pm}^{1\beta} R_{S^\pm}^{(5+j)\gamma} \nonumber\\
&&\hspace{1.4cm}
-\frac{1}{\sqrt{2}} \Big[  {Y_{\nu_{ji}}}  (\upsilon_u R_S^{(2+j)\alpha}+\upsilon_{\nu_j} R_S^{2\alpha}) -\lambda_i(\upsilon_d R_S^{2\alpha}+\upsilon_u R_S^{1\alpha})   \nonumber\\
&&\hspace{1.4cm}
 +2\kappa_{ijk}\upsilon_{\nu_j} R_S^{(2+k)\alpha} \Big] ({Y_{\nu_{li}}}R_{S^\pm}^{(2+l)\beta}R_{S^\pm}^{2\gamma} -\lambda_i R_{S^\pm}^{1\beta}R_{S^\pm}^{2\gamma}) \nonumber\\
&&\hspace{1.4cm}
+\frac{1}{\sqrt{2}}(A_\nu Y_\nu)_{ij} R_S^{(5+i)\alpha}R_{S^\pm}^{2\beta}R_{S^\pm}^{(2+j)\gamma} + \frac{1}{\sqrt{2}}(A_\lambda \lambda)_i R_S^{(5+i)\alpha}R_{S^\pm}^{2\beta}R_{S^\pm}^{1\gamma}\nonumber\\
&&\hspace{1.4cm}
+\frac{1}{\sqrt{2}}(A_e Y_e)_{ij} \Big[ R_S^{(2+i)\alpha}R_{S^\pm}^{(5+j)\beta}R_{S^\pm}^{1\gamma} -R_S^{1\alpha}R_{S^\pm}^{(5+j)\beta}R_{S^\pm}^{(2+i)\gamma} \Big].
\end{eqnarray}
The unitary matrices $R_S$, $R_{S^\pm}$ (and $R_u$, $R_d$, $Z_+$, $Z_-$ below) can be found in Ref.~\cite{Zhang1}.

The couplings between CP-even neutral scalars and squarks are written as
\begin{eqnarray}
\mathcal{L}_{int} = \sum\limits_{\alpha=1}^8 \sum\limits_{I,J=1}^6 ( C_{\alpha IJ}^{U^\pm} {S_\alpha U_I^- U_J^+} + C_{\alpha IJ}^{D^\pm} {S_\alpha D_I^+ D_J^-}),
\end{eqnarray}
with
\begin{eqnarray}
&&C_{\alpha IJ}^{U^\pm} = \frac{-e^2}{6\sqrt{2}}(\upsilon_d R_S^{1\alpha}-\upsilon_u R_S^{2\alpha}-\upsilon_{\nu_j} R_S^{(2+j)\alpha}) \Big[ \frac{4}{c_{_W}^2}\delta^{IJ} - \frac{3+2{s_{_W}^2}}{s_{_W}^2{c_{_W}^2}}R_u^{iI\ast}R_u^{iJ } \Big]  \nonumber\\
&&\hspace{1.4cm}
-\frac{1}{\sqrt{2}}
\Big[(A_u Y_u)_{ij}R_S^{2\alpha} - Y_{u_{ij}}\lambda_k (\upsilon_d R_S^{(5+k)\alpha}+\upsilon_{\nu_k^c} R_S^{1\alpha}) \nonumber\\
&&\hspace{1.4cm}
+ Y_{u_{ij}}Y_{\nu_{kl}} (\upsilon_{\nu_k} R_S^{(5+l)\alpha}+\upsilon_{\nu_l^c} R_S^{(2+k)\alpha}) \Big] (R_u^{iI\ast}R_u^{(3+j)J }+R_u^{(3+i)I\ast}R_u^{jJ })  \nonumber\\
&&\hspace{1.4cm}
-\sqrt{2}\upsilon_u Y_{u_{ik}}Y_{u_{jk}}R_S^{2\alpha}(R_u^{iI\ast}R_u^{jJ }+R_u^{(3+i)I\ast}R_u^{(3+j)J }),  \\
&&C_{\alpha IJ}^{D^\pm} = \frac{e^2}{6\sqrt{2}}(\upsilon_d R_S^{1\alpha}-\upsilon_u R_S^{2\alpha}-\upsilon_{\nu_j} R_S^{(2+j)\alpha})\Big[ \frac{2}{c_{_W}^2}\delta^{IJ} - \frac{1+2{s_{_W}^2}}{s_{_W}^2{c_{_W}^2}}R_u^{iI\ast}R_u^{iJ } \Big]  \nonumber\\
&&\hspace{1.4cm}
-\frac{1}{\sqrt{2}}(A_d Y_d)_{ij}R_S^{1\alpha} (R_d^{iI\ast}R_d^{(3+j)J }+R_d^{(3+i)I\ast}R_d^{jJ })  \nonumber\\
&&\hspace{1.4cm}
+\frac{1}{\sqrt{2}}Y_{d_{ij}}\lambda_k (\upsilon_u R_S^{(5+k)\alpha}+\upsilon_{\nu_k^c} R_S^{2\alpha}) (R_d^{iI\ast}R_d^{(3+j)J }+R_d^{(3+i)I\ast}R_d^{jJ })  \nonumber\\
&&\hspace{1.4cm}
-\sqrt{2}\upsilon_d Y_{d_{ik}}Y_{d_{jk}}R_S^{1\alpha}(R_d^{iI\ast}R_d^{jJ }+R_d^{(3+i)I\ast}R_d^{(3+j)J }).
\end{eqnarray}

The interaction Lagrangian between CP-even neutral scalars and charginos is formulated as
\begin{eqnarray}
\mathcal{L}_{int} = \sum\limits_{\alpha=1}^8 \sum\limits_{\beta,\gamma=1}^2 S_\alpha \bar{\chi}_\beta^+ \Big(C_{\alpha\beta\gamma}^{\chi^\pm}{P_L} + [C_{\alpha\gamma\beta}^{\chi^\pm}]^\ast{P_R}\Big) \chi_\gamma^- ,
\end{eqnarray}
where
\begin{eqnarray}
&&C_{\alpha\beta\gamma}^{\chi^\pm} =  \frac{-e}{{{\sqrt{2}s_{_W}}}}\Big[ R_S^{1\alpha }Z_+^{1\beta}Z_-^{2\gamma} + R_S^{2\alpha }Z_+^{2\beta}Z_-^{1\gamma} + R_S^{(2 + i)\alpha}Z_+^{1\beta}Z_-^{(2 + i)\gamma}\Big] \nonumber\\
&&\hspace{1.4cm}
- \frac{ {Y_{e_{ij}}}}{\sqrt{2}}\Big[ R_S^{1\alpha }Z_+^{(2+i)\beta}Z_-^{(2+j)\gamma} -R_S^{(2+i)\alpha }Z_+^{(2+j)\beta}Z_-^{1\gamma} \Big] \nonumber\\
&&\hspace{1.4cm}
-\frac{{Y_{\nu_{ij}}}}{\sqrt{2}}R_S^{(5+i)\alpha }Z_+^{2\beta}Z_-^{(2+j)\gamma}-\frac{{\lambda_i}}{\sqrt{2}}R_S^{(5+i)\alpha }Z_+^{2\beta}Z_-^{2\gamma},
\end{eqnarray}
and
\begin{eqnarray}
P_L=\frac{1}{2}{(1 - {\gamma _5})},\qquad  P_R=\frac{1}{2}{(1 + {\gamma _5})}.
\end{eqnarray}

\section{Form factors\label{app-form}}

The form factors are
\begin{eqnarray}
&&A_0(x)=-(x-g(x))/x^2, \\
&&A_{1/2}(x)=2\Big[x+(x-1)g(x)\Big]/x^2, \\
&&A_1(x)=-\Big[2x^2+3x+3(2x-1)g(x)\Big]/x^2,
\end{eqnarray}
with
\begin{eqnarray}
&&g(x)=\left\{\begin{array}{l}\arcsin^2\sqrt{x}, \hspace{2.6cm} x\le1;   \\
-{1\over4}\Big[\ln{1+\sqrt{1-1/x}\over1-\sqrt{1-1/x}}-i\pi\Big]^2, \quad x>1, \end{array}\right.
\end{eqnarray}
and
\begin{eqnarray}
&&F(x)=-(1-x^2)\Big({47\over2}x^2-{13\over2}+{1\over x^2}\Big)-3(1-6x^2+4x^4)\ln x
\nonumber\\
&&\hspace{1.45cm}
+{3(1-8x^2+20x^4)\over\sqrt{4x^2-1}}\cos^{-1}\Big({3x^2-1\over2x^3}\Big).
\end{eqnarray}

\section{Minimisation of the potential\label{app-mini}}

In the basis ${S'^T} = ({h_d},{h_u},{(\tilde \nu_i)^\Re},{({\tilde \nu_i^c})^\Re})$, the tree-level neutral scalar potential contains the following linear terms~\cite{mnSSM1}
\begin{eqnarray}
V_{\rm{linear}}^0 = t_{h_d}^0 {h_d}+t_{h_u}^0 {h_u}+ t_{(\tilde \nu_i)^\Re}^0 {(\tilde \nu_i)^\Re}+ t_{({\tilde \nu_i^c})^\Re}^0 {({\tilde \nu_i^c})^\Re},
\end{eqnarray}
where the different $t^0$ are the tadpoles at tree-level. They are equal to zero at the minimum
of the tree-level potential, and are given by
\begin{eqnarray}
\label{potential-1}
&&t_{h_d}^0=m_{{H_d}}^2 \upsilon_d + \frac{G^2}{4}( \upsilon_d^2 - \upsilon_u^2 +\upsilon_{\nu_i}\upsilon_{\nu_i}) \upsilon_d  - (A_\lambda \lambda)_i {\upsilon_u} \upsilon_{\nu_i^c} - {\lambda _j}{\kappa _{ijk}}{\upsilon_u}\upsilon_{\nu_i^c} \upsilon_{\nu_k^c}  \nonumber\\
&&\qquad\;\;  + \:({\lambda _i}{\lambda _j}\upsilon_{\nu_i^c}\upsilon_{\nu_j^c}  + {\lambda _i}{\lambda _i}\upsilon_u^2){\upsilon_d}  - {Y_{{\nu_{ij}}}}\upsilon_{\nu_i}({\lambda _k}\upsilon_{\nu_k^c}\upsilon_{\nu_j^c} + {\lambda _j}\upsilon_u^2),\\
\label{potential-2}
&&t_{h_u}^0=m_{{H_u}}^2{\upsilon_u} - \frac{{G^2}}{4}(\upsilon_d^2 - \upsilon_u^2 +\upsilon_{\nu_i}\upsilon_{\nu_i})\upsilon_u  + {(A_\nu Y_\nu)}_{ij}\upsilon_{\nu_i}\upsilon_{\nu_j^c} - (A_\lambda \lambda)_i {\upsilon_d} \upsilon_{\nu_i^c}  \nonumber\\
&&\qquad\;\;  + \:({\lambda _i}{\lambda _j}\upsilon_{\nu_i^c}\upsilon_{\nu_j^c}  + {\lambda _i}{\lambda _i}\upsilon_u^2){\upsilon_u} + {Y_{{\nu_{ij}}}}\upsilon_{\nu_i}({\kappa _{ljk}}\upsilon_{\nu_l^c} \upsilon_{\nu_k^c} - 2 {\lambda _j}\upsilon_d \upsilon_u )  \nonumber\\
&&\qquad\;\; - \: {\lambda _j}{\kappa _{ijk}}{\upsilon_d}\upsilon_{\nu_i^c} \upsilon_{\nu_k^c} +  ( Y_{\nu_{ki}}Y_{\nu_{kj}}\upsilon_{\nu_i^c}\upsilon_{\nu_j^c}  + Y_{\nu_{ik}}Y_{\nu_{jk}}\upsilon_{\nu_i}\upsilon_{\nu_j} )  \upsilon_u  ,\\
\label{potential-3}
&&t_{(\tilde \nu_i)^\Re}^0=m_{{{\tilde L}_{ij}}}^2\upsilon_{\nu_j} + \frac{{G^2}}{4}(\upsilon_d^2 - \upsilon_u^2 +\upsilon_{\nu_j}\upsilon_{\nu_j})\upsilon_{\nu_i}   + {(A_\nu Y_\nu)}_{ij}\upsilon_u\upsilon_{\nu_j^c}   \quad \nonumber\\
&&\qquad\quad\:\: - \:  {Y_{{\nu_{ij}}}}{\lambda _k}\upsilon_{\nu_j^c} \upsilon_{\nu_k^c}  \upsilon_d -{Y_{{\nu_{ij}}}}{\lambda _j}\upsilon_u^2  \upsilon_d + {Y_{{\nu_{il}}}}{\kappa _{ljk}}\upsilon_u \upsilon_{\nu_j^c} \upsilon_{\nu_k^c}   \quad \nonumber\\
&&\qquad\quad\:\: + \: {Y_{{\nu_{ij}}}}{Y_{{\nu_{lk}}}}\upsilon_{\nu_l}\upsilon_{\nu_j^c} \upsilon_{\nu_k^c} + \: {Y_{{\nu_{ik}}}}{Y_{{\nu_{jk}}}}\upsilon_u^2\upsilon_{\nu_j} ,\\
\label{potential-4}
&&t_{({\tilde \nu_i^c})^\Re}^0=m_{\tilde \nu_{ij}^c}^2 \upsilon_{\nu_j^c} + {(A_\nu Y_\nu)}_{ji}\upsilon_{\nu_j}\upsilon_u - (A_\lambda \lambda)_i{\upsilon_d}{\upsilon_u} +{( A_\kappa \kappa)}_{ijk} \upsilon_{\nu_j^c} \upsilon_{\nu_k^c}  \nonumber\\
&&\qquad\quad\:\: + \: {\lambda _i}{\lambda _j}\upsilon_{\nu_j^c}(\upsilon_d^2  + \upsilon_u^2) + 2{\kappa _{lim}}{\kappa _{ljk}} \upsilon_{\nu_m^c} \upsilon_{\nu_j^c} \upsilon_{\nu_k^c}
-  2{\lambda _j}{\kappa _{ijk}}{\upsilon_d}{\upsilon_u}\upsilon_{\nu_k^c} \nonumber\\
&&\qquad\quad\:\: -\: {Y_{{\nu_{ji}}}}{\lambda _k}\upsilon_{\nu_j} \upsilon_{\nu_k^c}{\upsilon_d} - {Y_{{\nu_{kj}}}}{\lambda _i}\upsilon_{\nu_k} \upsilon_{\nu_j^c} {\upsilon_d}  +  2{Y_{{\nu_{jk}}}}{\kappa _{ikl}}{\upsilon_u}\upsilon_{\nu_j} \upsilon_{\nu_l^c} \nonumber\\
&&\qquad\quad\:\: + \: {Y_{{\nu_{ji}}}}{Y_{{\nu_{lk}}}}\upsilon_{\nu_j}\upsilon_{\nu_l}\upsilon_{\nu_k^c} + {Y_{{\nu_{ki}}}}{Y_{{\nu_{kj}}}}\upsilon_u^2\upsilon_{\nu_j^c} .
\end{eqnarray}

\end{document}